\documentclass[numberedappendix]{emulateapj}

\usepackage{amsmath}
\usepackage{epsfig}
\usepackage{color}

\begin{document}

\title{A High-Frequency Doppler Feature in the Power Spectra of Simulated GRMHD Black Hole Accretion Disks}
\author{Sarah Wellons\altaffilmark{1,3}, Yucong Zhu\altaffilmark{1}, Dimitrios Psaltis\altaffilmark{2}, Ramesh Narayan\altaffilmark{1}, Jeffrey E. McClintock\altaffilmark{1}}

\altaffiltext{1}{Harvard-Smithsonian Center for Astrophysics, 60 Garden Street, Cambridge, MA}
\altaffiltext{2}{Astronomy Department, University of Arizona, 933 North Cherry Avenue, Tucson, AZ}
\altaffiltext{3}{swellons@cfa.harvard.edu}

\begin{abstract}
Black hole binaries exhibit a wide range of variability phenomena, from large-scale state changes to broadband noise and quasi-periodic oscillations, but the physical nature of much of this variability is poorly understood.  We examine the variability properties of three GRMHD simulations of thin accretion disks around black holes of varying spin, producing light curves and power spectra as would be seen by observers.  We find that the simulated power spectra show a broad feature at high frequency, which increases in amplitude with the inclination of the observer.  We show that this high-frequency feature is a product of the Doppler effect and that its location is a function of the mass and spin of the black hole.  This Doppler feature demonstrates that power spectral properties of the accretion disk can be tied to, and potentially used to determine, physical properties of the black hole.
\end{abstract}

\section{Introduction}
\label{sec:intro}

Accreting black holes in binary systems are variable on many time scales.  In the long term, they are observed to undergo changes in state as the mass accretion rate changes, from a bright state dominated by soft thermal emission, to a low-luminosity state dominated by a hard power-law spectrum (see \citealt{R+M2006} for a review).  These large-scale changes are interpreted \citep{Esin+1997,DGK2007,Narayan+McClintock2008,Yuan+Narayan2014} as the inner part of the accretion disk switching from an optically thick and geometrically thin disk \citep{Shakura+Sunyaev1973,Novikov+Thorne1973} to an optically thin and geometrically thick disk, or ADAF \citep{Narayan+Yi1994,Narayan+Yi1995,Abram+1995}.

On shorter timescales, the power spectra of observed accreting black holes consist of broad band noise spanning a large range of frequencies between a low and high break frequency plus, in some cases, sharper, narrower features known as quasi-periodic oscillations (QPOs, see \citealt{VDK2006} for a review).  Both components can be well-described by a series of Lorentzians \citep{BPK2002}, but the physical mechanisms by which the Lorentzian components are produced are poorly understood.  

Models for the broad band noise require a mechanism which produces variability over a large range of dynamical time- and length-scales.  The magneto-rotational instability (MRI, \citealt{Balbus+Hawley1991}), for example, creates turbulence in the accretion flow and is believed to be responsible for angular momentum transport in accretion disks.  This MHD turbulence does indeed generate fluctuations over a broad range of timescales \citep{Krolik+Hawley2002} and thus could potentially produce the sort of broad-band, power-law power spectrum that is typically observed.  Proposed mechanisms for QPOs are many and varied, including (but not limited to) tilted accretion flows \citep{Stella+Vietri1998, Markovic+Lamb1998,Ingram+Done2011}, diskoseismic oscillations \citep{Nowak+Wagoner1993, Kato2001, Silbergleit+2001} and resonances in the accretion disk \citep{Abram+2003}.

Identifying features in the power spectrum of an accreting black hole could characterize special frequencies in the disk related to the black-hole mass or spin.  For example, for a black hole of known mass, the identification of a power spectral feature with the orbital frequency at the innermost stable circular orbit (ISCO) of the accretion disk would indicate the black-hole spin $a_*$, since disks around black holes with higher spin have smaller ISCOs.

\begin{deluxetable*}{cccccccccc}
\tabletypesize{\footnotesize}
\tablecolumns{10}
\tablecaption{ GRMHD simulation parameters \label{tab:sims}}
\tablehead{
\colhead{$a_*$} & \colhead{$n_r$} & \colhead{$n_\theta$} & \colhead{$n_\phi$} & \colhead{$\dot{M}$ ($\dot{M}_{\rm Edd}$)} & \colhead{Length\tablenotemark{a} ($M$)} & \colhead{$f_{\rm QPO}~(1/M)$} & \colhead{$f_{\rm peak}\tablenotemark{b}~(1/M)$} & \colhead{$r_{\rm peak}\tablenotemark{c}~ (M)$} & \colhead{$r_{\rm isco}~(M)$}} \\
\startdata
0 & 256 & 64 & 64 & 0.7 & 40000 & $8 \times 10^{-4}$ & $9.5 \times 10^{-3}$ & 6.53  & 6.00 \\
0.7 & 256 & 64 & 96 & 0.64 & 20000 & $5 \times 10^{-4}$ & $1.45 \times 10^{-2}$ & 4.68 & 3.39 \\
0.9 & 256 & 64 & 96 & 0.72 & 30000 & $3 \times 10^{-4}$ & $1.57 \times 10^{-2}$ & 4.4 & 2.32
\enddata
\tablecomments{The dimensions of the simulation grid are given by the number of cells in each direction $n_r$, $n_\theta$, and $n_\phi$.  All simulations have $h/r\approx0.1$.}
\tablenotetext{a}{Length refers to the length of time in the simulation which is used for analysis, and does not include the first 20000 M where the simulation has not reached inflow equilibrium.}
\tablenotetext{b}{$f_{\rm peak}$ refers to the average peak frequency of $P_{\rm Dopp}$, calculated as described in Section \ref{sec:results}.}
\tablenotetext{c}{$r_{\rm peak}$ is the radius whose orbital frequency corresponds to $f_{\rm peak}$.}
\end{deluxetable*}

By examining the variability properties of simulations of accretion disks around stellar-mass black holes, we aim to shed light on the physical origins of some of these power spectral features and provide some tools for understanding observations.  In this paper, we analyze general relativistic magnetohydrodynamic (GRMHD) disk simulations from an observer's point of view.  We examine how the variability properties change with the observer's inclination and choice of photon energy band and identify power spectral features that may hint at black hole properties.  In Section \ref{sec:methods}, we describe the simulations themselves as well as the methods by which we produce the observed lightcurves and power spectra.  In Section \ref{sec:results}, we investigate a high-frequency feature in the power spectrum, determine its origin, and show how it changes with spin, inclination, and energy.  We discuss the implications for observers who seek to characterize black holes in Section \ref{sec:discuss}, and conclude in Section \ref{sec:conclusion}.  

Throughout, we use standard gravitational units and set $G=c=1$.  Thus, we express lengths in units of the black hole mass $M$, which corresponds to a physical unit of $GM/c^2$.  Similarly, our unit of time is also $M$, which corresponds to a physical unit of $GM/c^3$.

\vspace{1.5cm}
\section{Methodology}
\label{sec:methods}

\subsection{GRMHD Simulations}
\label{ssec:sims}

In this paper, we analyze simulations of accretion disks around black holes with three different spin parameters: $a_* \equiv a/M = (0,~0.7~,0.9)$.  The simulations are performed using the 3D GRMHD code HARM \citep{Gammie+2003,McKinney2006,McKinney+Blandford2009} in a Kerr spacetime with the appropriate value of $a_*$.  These are the same simulations analyzed by \citet{Zhu+2012} and \citet{
Kulkarni+2011}, but have been run longer.  Simulation length and other parameters are shown in Table \ref{tab:sims}.

Each simulation is initialized with an orbiting torus of gas in hydrostatic equilibrium in the equatorial plane around the black hole.  The gas is evolved according to the GRMHD equations and is driven to steady-state accretion via MHD turbulence that emerges as a result of the MRI.  This initial part of the simulation before inflow equilibrium is reached is not considered in the analysis that follows.  Additionally, the simulations are only converged (at a steady $\dot{M} \approx 0.7~\dot{M}_{\rm Edd}$) out to 15-20 $M$, so we disregard the region outside the radius of convergence.

The code HARM does not include radiative transfer.  In the present work, we use an ad hoc cooling prescription \citep{Shafee+2008,Penna+2010} whereby we effectively set the thickness of the disk.  (Section \ref{ssec:emission} describes how we reconstruct the emergent spectrum.)  All three simulations discussed here are thin disks with $h/r \approx 0.1$.

\subsection{Raytracing}
\label{ssec:raytracing}

\begin{figure}[t]
  \centering
  \includegraphics[width=\columnwidth]{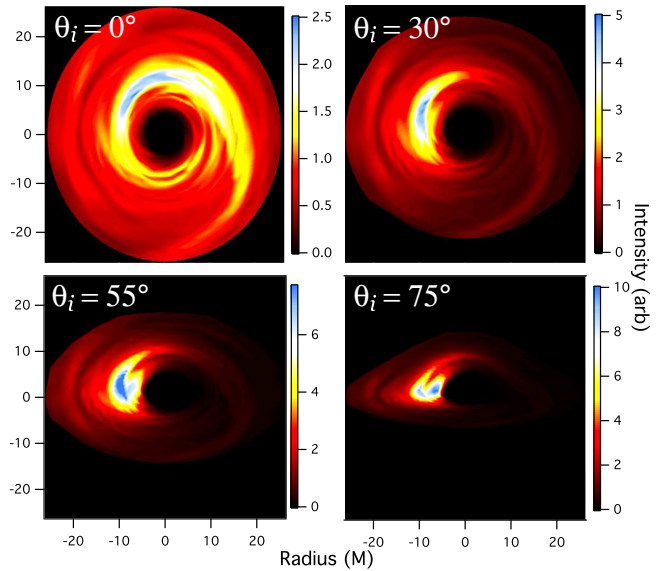}
  \caption{Raytraced image of a simulated accretion disk around a non-spinning black hole out to a radius $r$ = 25 M, at several inclinations.  Colors indicate bolometric intensity in arbitrary but consistent units.  As inclination increases, the image becomes warped as photon geodesics are deflected by the strong gravity near the black hole.  Also, the Doppler effect becomes sharper, boosting the left side of the disk where the orbital velocity points toward the observer.}
  \label{fig:imgincline}
\end{figure}

In order to produce the light curves that would be seen by a distant observer, we need to trace the paths taken by photons that are emitted at the disk midplane and navigate the curved spacetime around the black hole.  The strong gravity around the black hole deflects the paths of photons and produces a warped image for a distant observer, see Figure \ref{fig:imgincline} for examples.  To reproduce this, we construct an elliptical grid in the observer's image plane and follow the paths of photons back to their places of origin, as described by \citet{Kulkarni+2011}.  Initial conditions for the position vector of each ray are given by their location in the image plane, and initial conditions for the momentum vector are given by assuming that the observer is sufficiently far away from the black hole that all photons travel parallel to the line of sight.

The trajectory of the ray from each grid cell in the image plane is evolved along null geodesics in a Kerr spacetime until the ray strikes the plane of the disk or crosses the event horizon and is lost.  Then, knowing the location at which the ray struck (or emerged from) the disk as well as the strike angle, we reconstruct the emergent spectrum.

\subsection{Calculation of Emitted Spectrum}
\label{ssec:emission}

We use the stellar atmospheres code TLUSTY \citep{Hubeny+Lanz1995}, which has been adapted to the problem of black hole accretion disks (see \citealt{Davis+2005, Davis+Hubeny2006}), to calculate the radiation spectrum emerging from each disk annulus.  TLUSTY treats annuli in the accretion disk as one-dimensional plane-parallel atmospheres.  Under the assumption that the atmosphere is in hydrostatic equilibrium, the code solves for both the radiation field and the vertical structure of the annulus in the local comoving frame of the gas.

Free-free, free-bound, and bound-free radiative processes are all included, as well as an angle averaged Kompaneets treatment of Compton scattering \citep{Hubeny+2001}.  The physics of magnetic pressure support, convection, and heat conduction are, however, neglected.  Deviations from local thermodynamic equilibrium are allowed and the level populations for H, He, C, N, O, Ne, Mg, Si, S, Ar, Ca, and Fe are explicitly computed via relevant rate equations (under the assumption of solar abundances).  Only the lowest ionization energy level is considered for each ionization state, except for H and He where we include the lowest 9 and 4 levels respectively.  

With a final assumption that the local viscous heating is proportional to gas density, TLUSTY allows the full range of disk annuli to be encapsulated with just three parameters \citep{Davis+Hubeny2006}: (i) the vertical column mass $\Sigma$, (ii) the vertical tidal gravity $Q = g_\bot/z$ where $g_\bot$ is the vertical component of the gravitational acceleration and $z$ is height above the midplane, and (iii) the total emerging flux $F=\sigma T_{\rm eff}^4$.  With these three parameters and knowledge of the angle at which the ray emerges from the disk, the emergent spectrum is fully determined.  Note that TLUSTY computes the emerging intensity as a function of the angle of the ray with respect to the surface normal so that limb darkening is calculated self-consistently.  (At very high inclinations where $\cos\theta_i \gg h/r$, however, this method would not be valid since the radii at which the ray intersects the midplane and the photosphere would differ significantly.)

Since the properties of each annulus are uniquely specified by only three parameters, we have constructed a large precomputed grid of annuli to span the full parameter range for stellar-mass black-hole accretion disks.  In this work, we make use of a grid of over 40,000 TLUSTY models that were developed in \citet{Zhu+2012}. The use of a precomputed grid provides a significant computational time advantage; the overall spectra of many different disk models can be quickly computed by interpolating models from this grid.  When the atmosphere is optically thick, the spectra produced by these annuli will resemble a blackbody (plus an extended non-Wien tail from non-saturated Comptonization), but we are not limited to cases of high optical depth - the code has the advantage of handling optically thin atmospheres as well.

The strike angle comes from the raytracing as described above and $Q$ is simply a function of $r$; the other parameters must be retrieved from the simulation.  Because the simulation is 3D, while the procedure described above requires a flat disk, $T_{\rm eff}$ and $\Sigma$ are collapsed to the midplane along $z$.  The effective temperature at a point on the 2D disk is calculated by finding the total flux emitted within a vertical column.  Similarly, the column density is calculated by summing up the mass in all the cells in that column; see \citet{Zhu+2012} for details.

Having calculated these parameters for each midplane cell in the simulation, we retrieve the spectrum emitted in the fluid frame at the location where each ray strikes the disk using the appropriate pre-computed TLUSTY model.  Finally, we calculate the observed spectrum by applying a redshift which is determined by the radius, strike angle, and orbital velocity as described by \citealt{Kulkarni+2011}.  This redshift accounts both for gravitational redshift and the Doppler shift.  

The effect of the Doppler shift is visible in Figure \ref{fig:imgincline}.  A highly-inclined observer for whom the disk is rotating counter-clockwise will see bright, Doppler-boosted emission on the left where a large component of the fluid velocity is directed toward the line of sight.  Conversely, on the right, emission is boosted away from the observer and the disk is dim.

We have tested the method described in Sections \ref{ssec:raytracing}-\ref{ssec:emission} and shown that it accurately recovers the emission produced by a Novikov-Thorne model disk; see Appendix \ref{app:NT} for details.

\begin{figure}[t]
  \centering
  \includegraphics[width=\columnwidth]{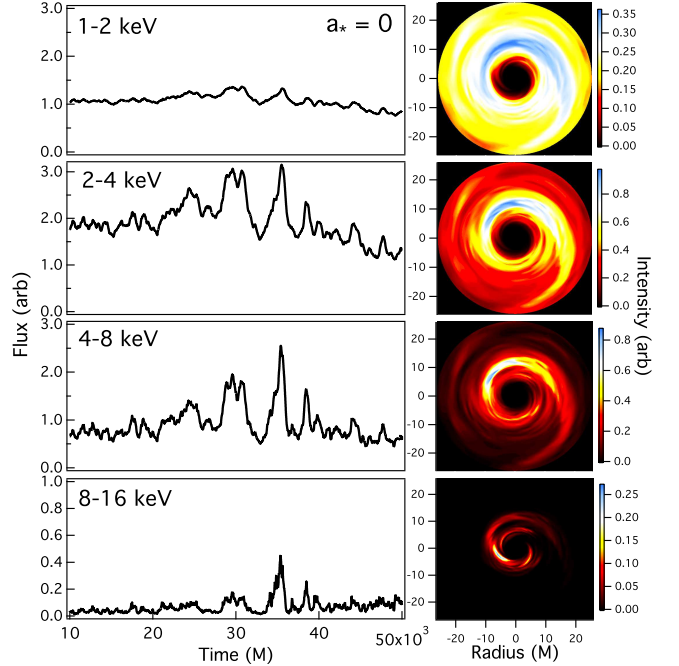}
  \caption{Light curves and images seen by an observer face-on to a simulated accretion disk around a non-spinning black hole out to a radius $r$ = 25 M, in several photon energy bands.  Units of flux and intensity are arbitary but consistent within this Figure and with Figure~\ref{fig:imgincline}.  As the band moves to higher photon energies, the lightcurves and images are dominated by fluctuations at small radii. }
  \label{fig:lcbands}
\end{figure}

\subsection{Light curves and power spectra}
\label{ssec:pspecs}

The observed flux at a particular time is calculated by integrating over the disk in the image plane using the spectra retrieved as described in Section \ref{ssec:emission}.  The light curves shown in this paper do not take into account the differences in light travel times between photons emitted at different locations on the disk.  Similar to \citet{Dexter+2010}, we found that including the delays has a negligible effect on the light curves and power spectra for a 10 $M_\odot$ black hole, so we ignore them to make the computation more straightforward.

Figure \ref{fig:lcbands} shows sample light curves at different photon energy bands.  A band which is in the Rayleigh-Jeans regime for the temperatures in the disk (such as the 1-2 keV panels at the top) varies smoothly over a broad range of radii, whereas a higher-energy band in the Wien regime (lower panels) is dominated by fluctuations in the hottest, innermost radii.  

At high energies, the spectral energy distribution is steep, so a small change in temperature results in a large change in intensity; this causes much larger variability in the light curves.

The primary tool used to characterize variability in X-ray binary systems is the power spectrum
\begin{equation}  P(f) = |F(f)|^2, \end{equation} 
where $F(f)$ is the Fourier transform of the lightcurve.  Following this practice, we use a Fast Fourier Transform method to compute the discrete Fourier transform and power spectrum of our light curves, which have temporal resolution $\Delta t$ = 5 M (the output cadence of the simulation, which corresponds to $\Delta t~\approx$ 250 $\mu$s for a 10 M$_\odot$ black hole).  We are able to substantially reduce the noise in the power spectrum by segmenting the light curves into $N_s$ = 2-4 pieces before taking the FFT and averaging the resulting power spectra together.  We then rebin the power spectra in frequency in such a way that each bin has equal size in $\log f$, averaging the power spectral components in each bin.  Then the error in the power spectrum is estimated to be
\begin{equation}  \sigma_S(f) = \frac{P(f)}{\sqrt{N_s N_{\rm bin}}}, \end{equation} 
where $N_{\rm bin}$ is the number of FFT bins contributing to each power spectrum frequency.  Typical values for $N_{\rm bin}$ within a single power spectrum range from a few at the low-frequency end to $\sim 200$ at the high-frequency end (due to the logarithmic spacing of the bins), such that the uncertainty in the calculated power spectrum ranges from  $\sim35\%$ to $3\%$.

\begin{figure}[t]
  \centering
  \includegraphics[width=0.95\columnwidth]{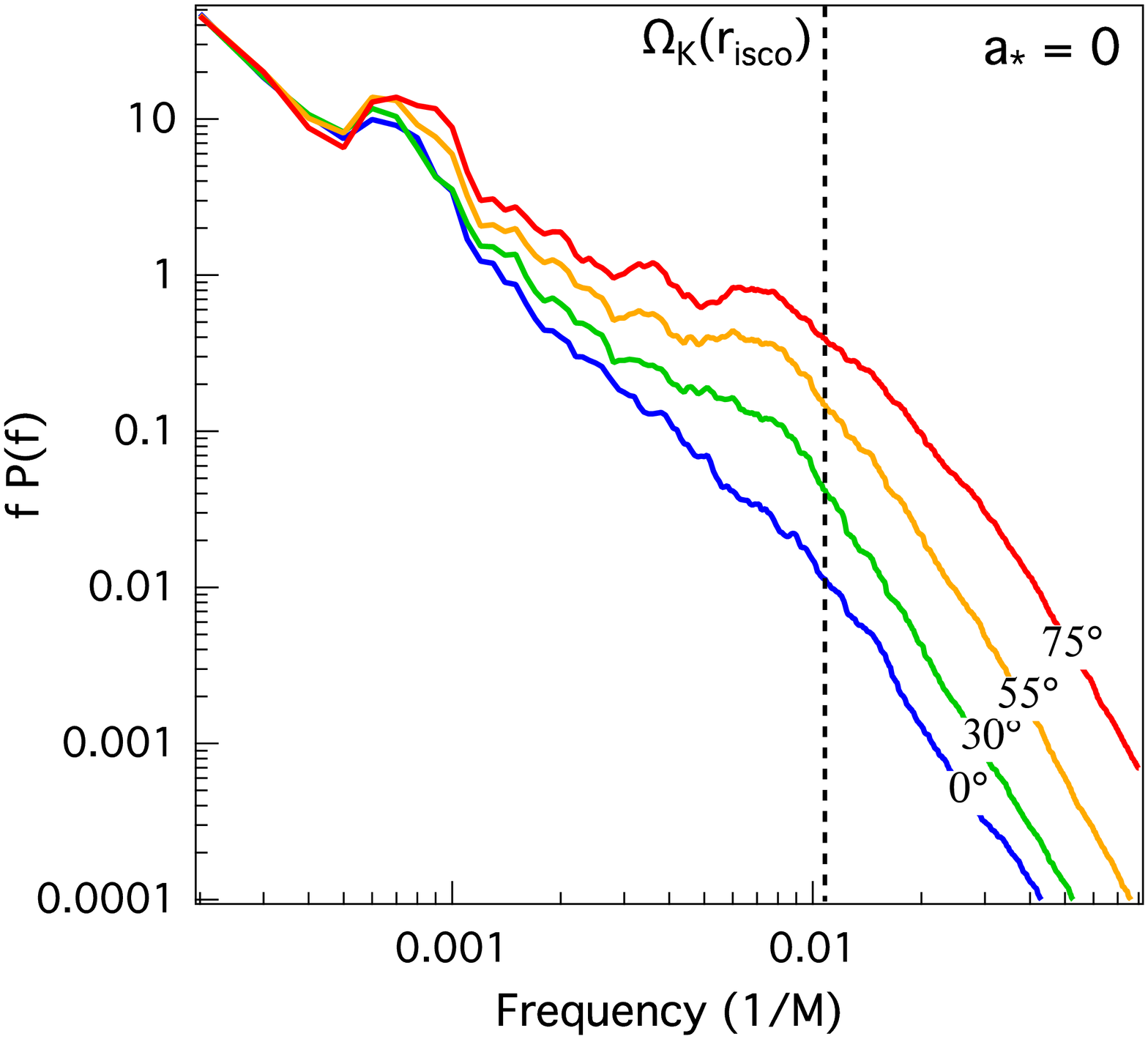}
  \caption{Power spectra of bolometric lightcurves seen by an observer at varying inclination to the accretion disk of a black hole with $a_*$ = 0.  The power spectra are well-described by one or more underlying power laws plus an apparent QPO around $10^{-3}/M$ and an additional broad feature from $(10^{-3}-10^{-2})/M$ which is stronger at higher inclinations.  For a 10$M_\odot$ black hole, a frequency of $10^{-2}/M$ corresponds to about 200 Hz.  The orbital frequency at the ISCO is plotted as a vertical dashed line; note that the broad feature is centered at a lower frequency, which suggests that the power originates at radii outside the ISCO.}
  \label{fig:pspecincl}
\end{figure}

In addition, we suppress leakage between frequencies by applying a Welch window to each segment,
\begin{equation}  w_k(t) = 1-\left(\frac{k-N/2}{N/2}\right)^2, \end{equation} 
where $N$ is the number of points in each segment and $k$ is the time index.  Because the Welch window de-weights the points at the ends of the segments, it effectively removes the information they carry from the power spectrum. To compensate for this, we take an additional $N_s-1$ segments which are offset from the original segments by half their length, and average these into the power spectra as well so that no information is lost \citep{NumericalRecipes}.  Examples of the resulting power spectra are shown in Figure \ref{fig:pspecincl}.

\section{Results}
\label{sec:results}

\subsection{Simulation Power Spectra}
\label{ssec:simspec}

The power spectra we obtain from our simulated lightcurves are characterized by three components, which can be seen for the non-spinning case in Figure \ref{fig:pspecincl}:
\begin{itemize}
\item[(i)] An underlying power law.
\item[(ii)] A QPO at low frequencies, $\lesssim 10^{-3}/M$ ($\sim$10 Hz for a 10 $M_\odot$ black hole).
\item[(iii)] A broad feature contributing extra power at higher frequencies, breaking off around $10^{-2}/M$ ($\sim$200 Hz for a 10 $M_\odot$ black hole).
\end{itemize}

As discussed in Section \ref{sec:intro}, the underlying power law $P(f) \propto f^{-\alpha}$ may be representative of the constant fluctuations caused by MHD turbulence associated with the MRI.  The power law index from the simulations ($\alpha \approx 3$) is steeper than typically observed or predicted ($\alpha = 2$), and the process producing it remains to to be precisely characterized.  (For a discussion of several possible mathematical models for the underlying variability, see \citet{PKC2008}, who also find a high-frequency feature similar to ours.)

The QPO is present for all spins and at all inclinations.  It also remains in approximately the same location throughout, though it moves slightly with spin from $\sim$16 Hz for $a_*$ = 0 to $\sim$10 Hz for $a_*$ = 0.7 and $\sim$6 Hz for $a_*$ = 0.9.  In this paper we focus on understanding the origin of the high-frequency feature and defer discussion of the QPO to a future paper.

The effect on the broad high-frequency feature of changing the observer's inclination is evident in Figure \ref{fig:pspecincl}.  As inclination increases, the feature becomes more and more prominent.  These changes with inclination strongly suggest that the Doppler effect is of great importance for this feature - as the observer becomes more inclined with respect to the disk, the component of orbital motion along the line of sight increases, strengthening the boost supplied by the Doppler effect.  As a result, at high inclinations the radii closest to the black hole possessing the highest orbital velocities are emphasized.

\begin{figure}[t]
  \centering
  \includegraphics[width=\columnwidth]{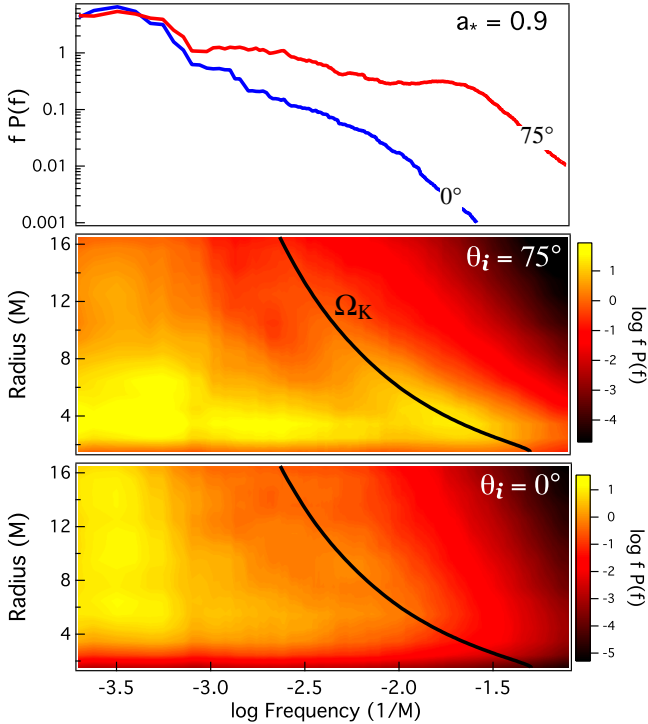}
  \caption{Top panel: Power spectra of bolometric light curves seen by observers who are face-on ($0^\circ$, blue) and at high inclination ($75^\circ$, red) to the accretion disk of a black hole with $a_*$ = 0.9.  Bottom panels: The same power spectra, broken down by radius of origin of the photons.  At high inclination, each annulus contributes additional power at its orbital frequency (black line), which results in the broad high-frequency feature in the combined power spectrum.}
  \label{fig:2dpspec}
\end{figure}

An examination of the power spectra broken down by radius confirms this emphasis on smaller radii.  As seen in Figure \ref{fig:2dpspec}, at high inclination each annulus contributes additional power at its own orbital frequency (the black line on the bottom two panels).  This excess is not present for a face-on observer.  Physically, this can be understood to mean that there are structures in the accretion disk which survive for at least one orbital period.  As a feature passes through the left side of the disk it experiences a Doppler boost and appears brighter, then dims as it passes on the right, and brightens again as it comes back around, repeating this process as long as the structure is present and thereby producing periodicity at the orbital frequency.  Based on this explanation, we refer to the resulting feature in the power spectrum as the ``Doppler feature."

\begin{figure}[t]
  \centering
  \includegraphics[width=\columnwidth]{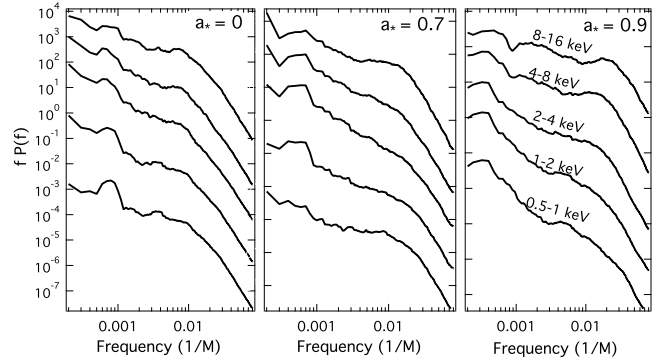}
  \caption{Power spectra of light curves seen by an observer at 55 degrees inclination in several energy bands for each of the three spins.  As the band moves to higher energies, the high-frequency Doppler feature becomes more pronounced.  Despite its origin in the thermal disk, the Doppler feature is noticeably present at high energies due to the steepness of the high-energy tail.  (The different curves in each panel have been shifted vertically for clarity.)}
  \label{fig:pspecband}
\end{figure}

Because the orbital frequency changes as a function of radius, the Doppler feature is necessarily broad, extending up to the orbital frequency of the radius which dominates the variability.  This radius does not appear to change with inclination or the energy band in which the disk is observed (except perhaps at the highest spin), though these two quantities do certainly affect how pronounced the feature is.  As can be seen in Figure \ref{fig:pspecband}, the Doppler feature is present at all energies, both soft and hard.  Moving to higher energies increases the prominence of the feature, since the spectrum is steep at those energies and a small change in temperature produces a larger change in intensity than in the lower-energy bands where the spectrum is shallower.

\begin{figure*}[t]
  \centering
  \includegraphics[width=1.5\columnwidth]{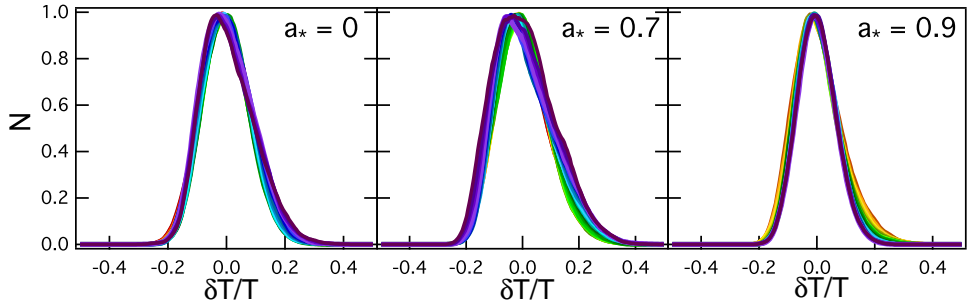}
  \caption{Distributions of temperature within an annulus from the simulations, for annuli spanning 2-10 M for each of the three spins.  The distributions are normalized against and centered on the median temperature of each annulus.  The distributions maintain a similar profile over all radii, with a characteristic half-width half-max $\delta T/T \sim 0.1$.  Different annuli are represented by different colors.}
  \label{fig:dT}
\end{figure*}

The location of the Doppler feature does change with spin, however, moving to higher frequencies as spin increases (see Table \ref{tab:sims}).  At higher spins, circular orbits are stable closer to the black hole, so the inner edge of the accretion disk reaches smaller, hotter, and faster-moving radii.  Accordingly, the higher orbital frequencies of these radii are visible in the power spectrum.  A priori, one might expect the Doppler feature to be dominated by the orbital frequency of the ISCO, but this is not the case.  In fact, the feature tends to peak or break at frequencies corresponding to radii up to $\sim2$ M \emph{outside} the ISCO, implying that there is another location in the disk which is dominant, and necessitating a more detailed explanation.  

The location of the break is affected not only by the strength of the Doppler effect, which increases towards smaller radii, but also by the emissivity of the disk, which peaks further out.  In the following sections, we model the shape of the Doppler feature as a function of spin, inclination, and temperature profile, and show how its break lies between the orbital frequencies at the ISCO and at the peak of emissivity.

\vspace{5mm}
\subsection{A Simplified Model of the Doppler Feature}
\label{ssec:model}

As a hotspot travels around the disk, its intensity as seen by an inclined observer is modulated by the Doppler effect.  Of course, for a smooth, non-fluctuating disk, this modulation will not produce any variability; some background fluctuations (such as those provided by the MHD turbulence) are required.  Thus the strength of the signal from the Doppler effect for a given radius will depend on (i) the strength of the underlying fluctuation, $\delta F(r)$, (ii) the orbital frequency $\Omega(r)$, and (iii) the observer's inclination.  The observer's inclination is known and the orbital frequency is easily determined analytically, but $\delta F(r)$ is not as obvious.

In the simulations, we have found that fluctuations in temperature within an annulus are approximately normally distributed, with a half-width half-max $\delta T/T \equiv \delta_T \sim 0.1$ which is consistent across all radii. (See Figure~\ref{fig:dT}.)  This means that the typical size of fluctuations in a given annulus scales with the typical flux at that radius.  Given an orbital frequency $\Omega(r)$ and temperature profile $T(r)$, we can therefore estimate the Doppler power as a function of radius for an observer at any inclination.

In the absence of any lensing effects (an approximation which is obviously crude, but which simplifies the analysis), the total flux for an observer at an angle $\theta_i$ to the disk is
\begin{equation} \label{eq:imgpl}  F(E) = \frac{\cos \theta_i}{D^2} \int r~dr\int I(E, r, \phi, t)~d\phi, \end{equation} 
where $D$ is the distance to the observer and $I(E, r, \phi, t)$ is the intensity emitted at the observer's photon energy $E$ at the location $(r, \phi)$ on the disk midplane at time $t$.  Since we have determined from the simulations that the typical scale of fluctuations is independent of radius, we treat $I$ as the sum of constant and time-varying parts
\begin{equation}  I(E, r, \phi, t) = I(E, r)[1 + \delta_I \cos(\phi - \Omega t)], \end{equation} 
where the second term represents a fluctuation of scale $\delta_I$ which orbits in the disk with frequency $\Omega(r)$.  

To apply a known disk emission profile $I_e$ (all subscripts $e$ indicate the frame of an observer comoving with the fluid at the midplane), we must convert between the frames of our stationary, distant observer and our comoving, midplane observer.  The photon energies in the two frames, $E$ and $E_e$, are related as follows,
\begin{equation} \frac{E_e}{E} = \frac{h\nu_e}{h\nu} = 1+z, \end{equation} 
where $h$ is the Planck constant, $\nu$ and $\nu_e$ are the frequencies of the photon in the two frames, and $1+z$ is the redshift factor.  Then since $I/\nu^3$ is invariant,
\begin{equation}  I(E,r) = I_e(E_e,r) \left(\frac{E}{E_e}\right)^3 = \frac{I_e[(1+z)E,r]}{(1+z)^3}. \end{equation} 
To return to the distant observer's energy $E$ we take a first-order Taylor expansion of the emission profile so that
\begin{equation} \label{eq:taylor} I(E,r) = \left[I_e(E,r) + \left. \frac{dI_e}{dE} \right|_{E} zE \right] \frac{1}{(1+z)^3}. \end{equation} 
The redshift factor $1+z$, which includes both gravitational redshift and Doppler shifts, is calculated for the Kerr metric as a function of $r$, $\phi$, $a_*$, and $\theta_i$ in Appendix \ref{sec:redshift}.

Combining Equations \eqref{eq:imgpl}-\eqref{eq:taylor}, we find that the time-varying component of the flux $F_{\rm var}$ changes with $r$ as
\begin{equation}
\frac{dF_{\rm var}}{dr} \propto r \int d \phi \left[I_e(E,r) + \left. \frac{dI_e}{dE} \right|_{E} zE \right] \frac{\cos[\phi - \Omega(r)t]}{(1+z)^3}.
\end{equation}
Given a choice of $I_e$, this integral is fully specified and may be evaluated numerically.  In the following analysis, we use a blackbody, $I_e(E,r) = B[E, T(r)],$ so that all that remains is a choice of temperature profile.  For direct comparison with simulation results, we use the temperature profiles recovered from the simulation, but any reasonable profile (e.g., a Novikov-Thorne profile) will provide similar results.  

The variable part of the flux $F_{\rm var}(t)$ at a radius $r$ follows a sinusoidal pattern as the fluctuation travels around the disk, providing power at the orbital frequency $\Omega(r)$, with the amount of power scaling with the amplitude of the sinusoid.  The relevant quantity for comparison with the power spectrum $\nu P(\nu)$ is the amplitude of $(dF_{\rm var}/d\ln\Omega)^2$, which can be derived from $dF_{\rm var}/dr$ as
\begin{equation}
\frac{dF_{\rm var}}{d\ln\Omega} = \Omega \frac{dF_{\rm var}}{dr} \frac{dr}{d\Omega}.
\end{equation} 

Finally, we note that in the simulation, the intrinsic variability present in the disk is more complicated than the simple cosine assumed above.  The fluctuations are not constant nor are they of a single size; rather, they have a range of sizes as shown in Figure \ref{fig:dT} and appear and decay over many timescales.  Thus, for a face-on observer, this intrinsic variability appears in the power spectrum not as a single delta function, but as a power law over a broad range of frequencies.  For inclined observers, this variability will be modulated (or multiplied) by the boosts supplied by the Doppler effect, and in Fourier space, this multiplication becomes a convolution.  Before making the comparison with simulations, therefore, we convolve our estimate of the Doppler power with a power law extending to higher frequencies to produce a final estimate of the shape of the Doppler feature $P_{\rm Dopp}$:
\begin{equation} P_{\rm Dopp} = \left( \frac{dF_{\rm var}}{d\ln\Omega}\right)^2 \otimes f^{-\alpha} \end{equation}
where $\alpha$ is the power law exponent at high frequencies that we fit to the simulated power spectrum for a face-on observer.

\begin{figure*}[t]
  \centering
  \includegraphics[width=1.7\columnwidth]{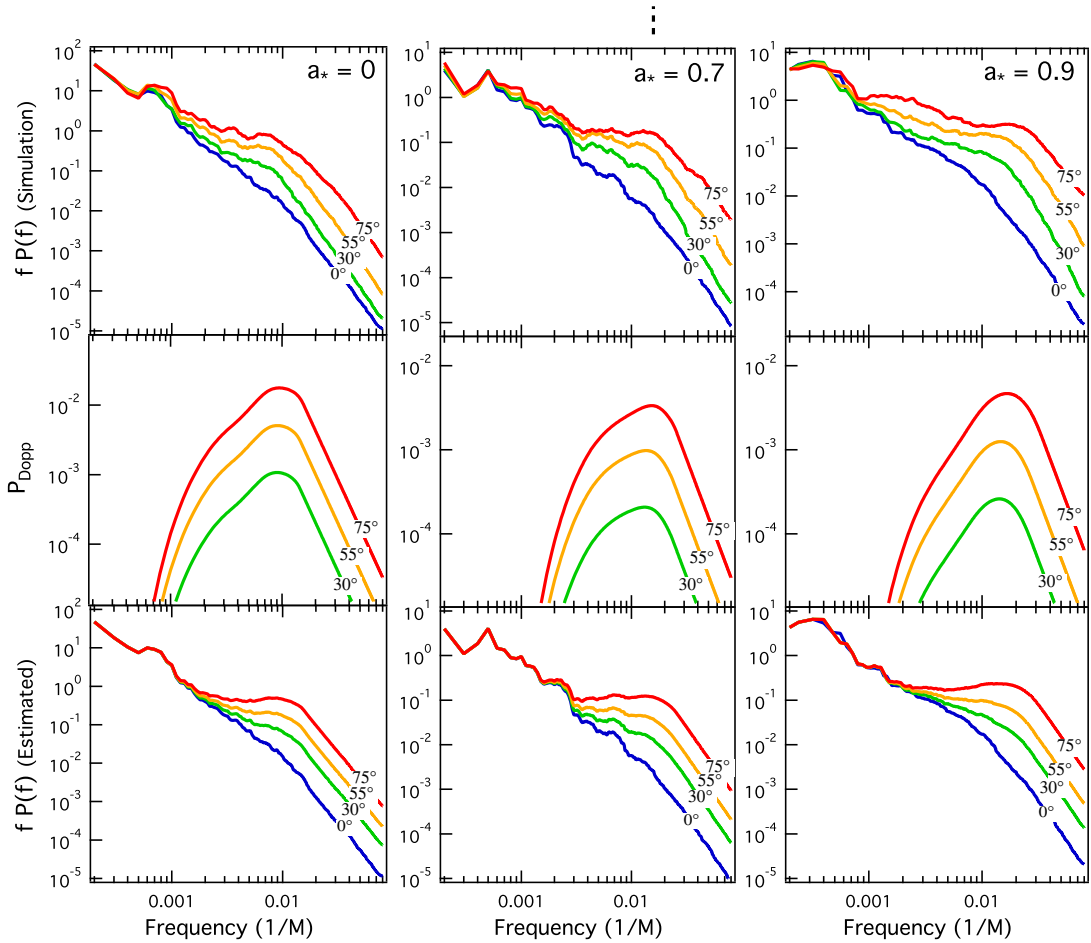}
  \caption{Upper panels: Bolometric power spectra from the simulations for each of the three spins, at several inclinations.  Middle panels: Estimate of the power from Doppler boosting of intrinsic fluctuations, $P_{\rm Dopp}$, for each of the spins and inclinations in the first row.  (For an observer at $0^\circ$ inclination, there is no Doppler power.)  Lower panels: Sum of the simulation power spectra at $0^\circ$ inclination and the estimated additional power from Doppler boosting for inclined observers.  Note the good agreement of these semi-analytic power spectra with the simulation results shown in the upper panels.}
  \label{fig:breakloc}
\end{figure*}

The result of this estimation can be seen in Figure \ref{fig:breakloc}.  In each column, the top panel shows the power spectrum calculated directly from the simulation.  In the bottom two sets of panels, we try to replicate the behavior of the power spectra analytically as described above - the central panels show our estimate of the Doppler feature $P_{\rm Dopp}$ while the bottom panels combine this estimate with the power spectrum for a face-on observer.  $P_{\rm Dopp}$ peaks at frequencies consistent with the breaks that appear in the simulations.  Because of the strong dependence on redshift $1+z$ (which is a function of inclination $\theta_i$) it is stronger at higher inclination, and at high spin the location of the peak changes slightly with inclination as well. The top and bottom sets of panels are in good agreement, indicating that we have identified the origin of the high-frequency feature.  In the following section, we discuss how this may apply to observations.

\section{Discussion}
\label{sec:discuss}

Observations of accreting black holes are typically sorted into ``states,'' determined by their X-ray flux and spectral hardness.  The canonical states are the ``thermal,'' ``hard,'' "quiescent," and ``steep power law'' states \citep{R+M2006}, though there exist many intermediate forms as well \citep{KW+VDK2008}.  Our simulated disks most closely correspond to accreting black holes in the thermal or SPL state, as they are dominated by thin accretion disks at high accretion rates ($\approx 70 \% ~\dot{M}_{\rm Edd}$) and thus high luminosities.  In all states, the power spectra can be described by the superposition of several Lorentzian components \citep{BPK2002}, bounded by a low-frequency component at one end and two high-frequency components on the other.  QPOs are represented as Lorentzians with a high quality factor $Q$ and are sharply peaked.

Because we have only 40000 M (2 seconds for a 10 M$_\odot$ black hole) of simulation data at best, we can only compare to the high-frequency end of the observed power spectra.  In particular, our highest-frequency feature ought to correspond to the highest-frequency component for observed black hole binaries.  We have seen in Section \ref{sec:results} that the power spectra from our simulated disks have a feature at high frequencies which is induced by the orbital motion of structures in the disk and highlighted by the Doppler effect.  The strength and location of this Doppler feature is affected by four quantities: the inclination of the observer, the energy band used by the observer, the black hole mass, and the black hole spin.

\begin{figure}[t]
  \centering
  \includegraphics[width=0.9\columnwidth]{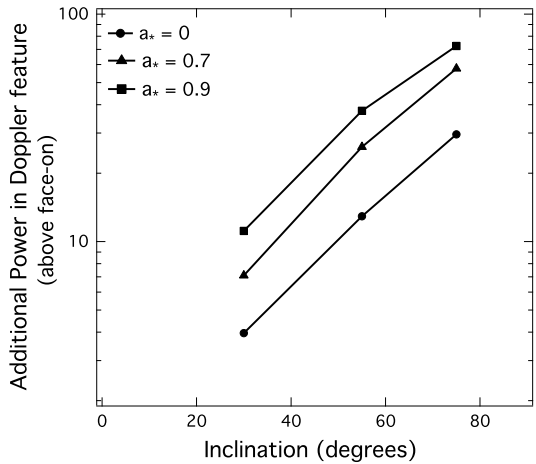}
  \caption{Additional power at the location of the peak of the Doppler feature for an inclined observer, normalized against the power for a face-on observer.  Observers at higher inclinations have a higher fraction of the orbital motion along the line of sight, and observe more variability power because of the Doppler effect.}
  \label{fig:extrapow}
\end{figure}

As discussed in Section \ref{sec:results}, the observer's inclination with respect to the disk as well as the energy band chosen by the observer each affect the strength of the Doppler feature in the power spectrum.  As the feature is a direct result of the Doppler effect, a more inclined observer has a larger component of the fluid velocity along the line of sight, a higher Doppler factor, and thus a stronger power spectral feature (Figure \ref{fig:extrapow}).  Moving to high energies also enhances the feature: Past the peak of the blackbody, the radiation spectrum becomes steeper so that the shifts in frequency provided by the Doppler effect produce a larger shift in intensity than in bands where the radiation spectrum is shallower (Figure \ref{fig:pspecband}).  

\begin{figure}[t]
  \centering
  \includegraphics[width=\columnwidth]{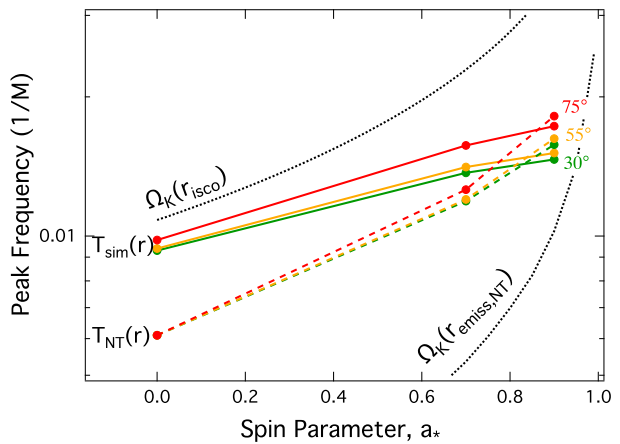}
  \caption{Frequency at which the estimated shape of the Doppler feature, $P_{\rm Dopp}$, peaks as a function of spin for several inclinations.  Solid lines correspond to the temperature profiles from the simulations, and dashed lines to Novikov-Thorne profiles \citep{Page+Thorne1974}.   Regardless of the specific $T(r)$, the peak shifts to higher frequencies as black hole spin increases.  The orbital frequencies at the ISCO and at the peak of $dL/d$ln$r$ (dotted lines) are plotted for comparison.  Note that the peak of $P_{\rm Dopp}$ always lies between the two, indicating that its location is a compromise between the strength of the Doppler effect and the luminosity profile of the disk.}
  \label{fig:spinbreak}
\end{figure}

We have shown in Figure \ref{fig:breakloc} that the location of the Doppler feature agrees with the peak of the function $P_{\rm Dopp}(\nu)$ derived in Section \ref{ssec:model}, which estimates the amplitude of Doppler modulations of hotspots produced by the temperature fluctuations present in the disk and their contribution to the power at orbital frequencies $\Omega(r)$.  The shape of this function depends on the spin of the black hole, $a_*$, as well as the temperature profile of the disk, $T(r)$.  At higher spins, the inner edge of the accretion disk extends closer to the black hole, providing access to the higher orbital frequencies at those radii and shifting the peak of $P_{\rm Dopp}$ to higher frequencies.  This effect is present regardless of the temperature profile chosen.  Figure \ref{fig:spinbreak} shows the location of the peak for the temperature profiles measured from the simulation $T_{\rm sim}(r)$ as well as an analytical $T_{\rm NT}(r)$ from \citet{Page+Thorne1974}, and both choices of temperature profile show the peak moving to higher frequencies with increasing black-hole spin as the accretion disk extends to smaller radii.

Observationally, the location of the highest-frequency Lorentzian component has been seen to shift to higher frequencies as accreting black holes undergo state transitions from low to high luminosity \citep{KW+VDK2008}, which is consistent with the idea that it represents some dominant radius that moves closer to the black hole as luminosity increases and the truncation radius marking the transition between the thin and thick disks moves inward.  At the highest luminosities where the accretion disk is expected to extend all the way to the ISCO, therefore, one could potentially use the location of this high-frequency feature to estimate spin in cases where the black hole mass is known.

One interesting implication of the results shown in Figure~\ref{fig:spinbreak} is that the variation of the Doppler feature frequency with black hole spin $a_*$, as derived from the simulations (solid lines), is fairly shallow. This is very different from the much steeper variation predicted by the Novikov-Thorne model (dashed lines) or the orbital frequency at the ISCO, $\Omega_K(r_{\rm isco})$. Models of high-frequency QPOs based on dynamical frequencies in the disk \citep{Psaltis2004} typically predict steep variations akin to the latter.  If the Doppler feature described in this paper is relevant to observed QPOs, then it is clear that simulation data will play an increasingly important role in interpreting observations.

The thin accretion disk simulations presented in this paper represent the state-of-the-art. Nevertheless, they omit some important physics. Notably, the code we used does not include radiative transfer. Instead, radiative cooling is handled via a toy prescription \citep{Shafee+2008, Penna+2010} which is designed to keep the disk geometrically thin. New codes are currently being developed that can handle radiation GRMHD \citep{Olek+2013, Olek+2014}. Simulations in the future with these codes will provide more reliable profiles of $T_{\rm sim}(r)$ and hence more accurate estimates of the Doppler feature frequency.

The origin of the QPO which appears in our power spectra at lower frequencies is as yet undetermined.  It moves to slightly lower frequency as the spin of the black hole increases, which is somewhat unexpected.  It may be that the QPO is an artifact of the simulation, e.g. some characteristic frequency of the torus with which the simulations are initialized, but this remains to be seen.  It is unlikely that the QPO corresponds to any real mode present in the disk, since the MRI tends to damp out these modes and there is no obvious mechanism to replenish them in the simulations. 

\section{Conclusion}
\label{sec:conclusion}

We have shown that our simulated GRMHD accretion disks have some of the same variability properties observed in black hole binaries.  These variability properties, in particular the feature observed at high frequencies, are a function of the physical properties of the system such as black-hole mass and spin.

The high-frequency feature originates from the orbital motion in the disk.  Hotspots which survive for at least one orbital period are alternately Doppler boosted and dimmed for an inclined observer, producing periodicity and hence additional power at orbital frequencies.  This Doppler feature peaks or breaks at the orbital frequency of the radius whose Doppler fluctuations are dominant, which is a function of spin (and, to a lesser degree, inclination).  This feature is stronger at higher inclinations, where the Doppler effect is stronger, and at higher energies past the peak of the blackbody spectrum, where the SED is steep.

Thus, given the disk temperature profile (e.g., from simulations), an observer could potentially use the location of the highest-frequency Lorentzian component in the power spectrum to determine the spin of an accreting black hole with known mass.  This procedure would be most plausible in the bright thermal or steep power law states, where the thin disk component is believed to extend all the way down to the ISCO.

While we were successful in reproducing some of the observed characteristics of accreting black holes and understanding the origin of the highest-frequency feature, our understanding of black hole binary variability is far from complete.  In particular, the origin of the low-frequency QPO which appears in our power spectra is still unknown.  It is not clear whether it is a consequence of some physical process or an artifact of the simulation.  Varying the parameters of the simulation itself (e.g., the location of the initial torus) may reveal whether it is a physical or computational effect.  Additionally, increasing the length of the simulations would provide more low-frequency information in the power spectra and help to better characterize this feature.

We will also aim to better describe these accreting black hole systems in the future by introducing more realistic physics into the simulations.  The observed profile of emission would be significantly improved by incorporating full radiative transfer into the simulations, rather than employing our current method of using an ad-hoc cooling prescription and doing radiative transfer in post-processing.  This step is crucial to fully understanding the radiative processes at work.  These goals are very ambitious, but would represent a significant improvement in our ability to faithfully simulate accreting black holes.

\acknowledgements

S.W. is supported by the National Science Foundation Graduate Research Fellowship under grant number DGE1144152.  Resources supporting this work were provided by the Extreme Science and Engineering Discovery Environment (XSEDE), which is supported by National Science Foundation grant number OCI-1053575, and the NASA High-End Computing (HEC) Program through the NASA Advanced Supercomputing (NAS) Division at Ames Research Center.  This material is based upon work supported by National Science Foundation grant number AST1312651 to R.N.  D.P. was partially supported by NSF CAREER award AST-0746549.

\appendix

\section{Test with Novikov-Thorne Disks}
\label{app:NT}

To test the raytracing code, we wish to see whether it can accurately recover all of the energy that is emitted at the disk midplane using distant observers scattered around the sky.  For an accretion disk with a Novikov-Thorne flux profile $F(r)$ \citep{Novikov+Thorne1973, Page+Thorne1974}, we know analytically that the luminosity per radius as observed at infinity is given by
\begin{equation}  \frac{dL}{dr} = 4 \pi r u_t(r) F(r), \end{equation} 
where the factor $u_t$, i.e. the covariant time component of the four-velocity, accounts for gravitational redshift.  The goal of this Appendix is to reproduce this analytical profile numerically by raytracing back to a Novikov-Thorne disk to retrieve the derivative $dL/dr$ at a particular location $(\theta, \phi)$ on the sky, then integrating over observers at all locations on the sky to retrieve the universal derivative $dL/dr$.  This is a comprehensive test of our raytracing code.

To compute the derivative $dL/dr$ for a single observer, we must integrate over the observer's image plane. In our raytracing code, the image plane is an elliptical grid in radial and angular coordinates ($b$, $\beta$) which is squeezed by a factor of $\cos\theta_i$.  The total flux in the image plane at that inclination is therefore
\begin{equation}  F \propto \cos\theta_i \int I(b,\beta)b~dbd\beta \propto \cos\theta_i \int I(b,\beta)~b^2~d(\log b)d\beta, \end{equation} 
where $I(b, \beta)$ is the intensity at infinity of the ray through the point with coordinates ($b,\beta$) as recovered from the accretion disk by raytracing.  Each angle $\beta$ has its own flux profile along the image plane radius $b$,
\begin{equation}  \frac{dF}{d\log b}(b,\beta) \propto \cos\theta_i ~I(b,\beta)~b^2~d\beta, \end{equation} 
or, converting between image plane radius $b$ and disk plane radius $r$,
\begin{equation}  \frac{dF}{dr}(r,\beta) \propto \cos\theta_i~I[b(r,\beta),\beta]~b(r)^2~\frac{d\log b}{dr}~d\beta. \end{equation} 
(Note that $d\log b/dr$ can be different for each $\beta$.)  Then the radial profile over the whole image plane, summed over all angles, is
\begin{equation}  \frac{dF}{dr}(r) \propto \cos\theta_i \sum\limits_\beta I[b(r,\beta),\beta]~b(r)^2~\frac{d\log b}{dr}. \end{equation} 
We calculate the derivative $d\log b/dr$ as a first-order numerical derivative along each angular coordinate $\beta$, and interpolate the $dF/dr(r,\beta)$ derivative to a common set of radii before summation.  

Then to integrate over the full $4 \pi$ sphere, we perform raytracing for observers at $\theta_i = [2.5^\circ, 7.5^\circ, 12.5^\circ,..., 90^\circ]$, so that the total luminosity of the disk is
\begin{equation}  L \propto \int F(\theta_i) \sin\theta_i d\theta_i d\phi \propto \sum\limits_{\theta_i} F(\theta_i) \sin\theta_i, \end{equation} 
where $F(\theta_i)$ is the flux observed at an inclination $\theta_i$.  For each of these inclinations, we find the flux profile $dF/dr$ as described above; then the total observed luminosity profile is
\begin{equation}  \frac{dL}{dr} \propto \sum\limits_{\theta_i} \frac{dF}{dr}(r, \theta_i)~ \sin\theta_i~. \end{equation} 

\begin{figure}[t]
  \plottwo{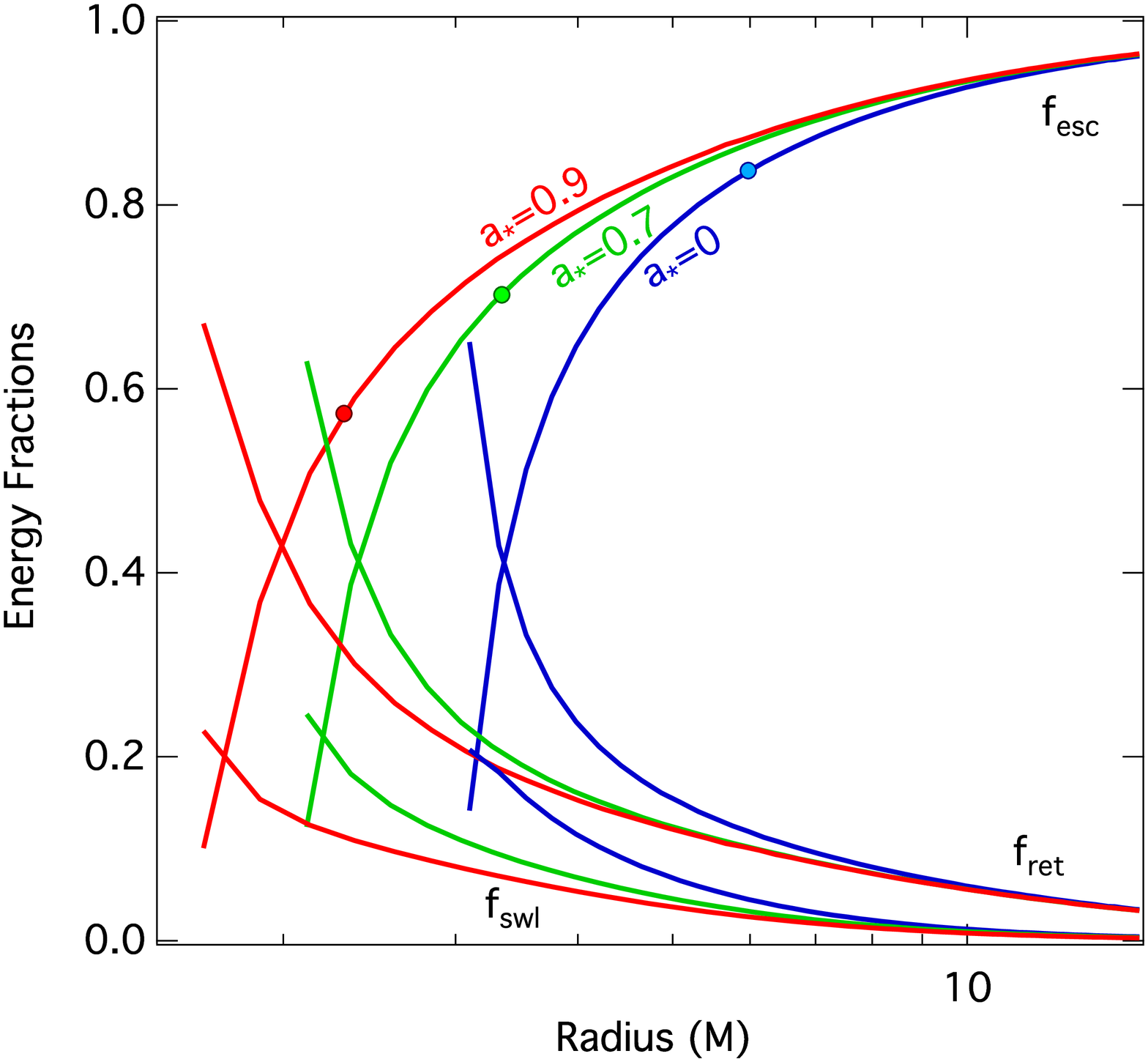}{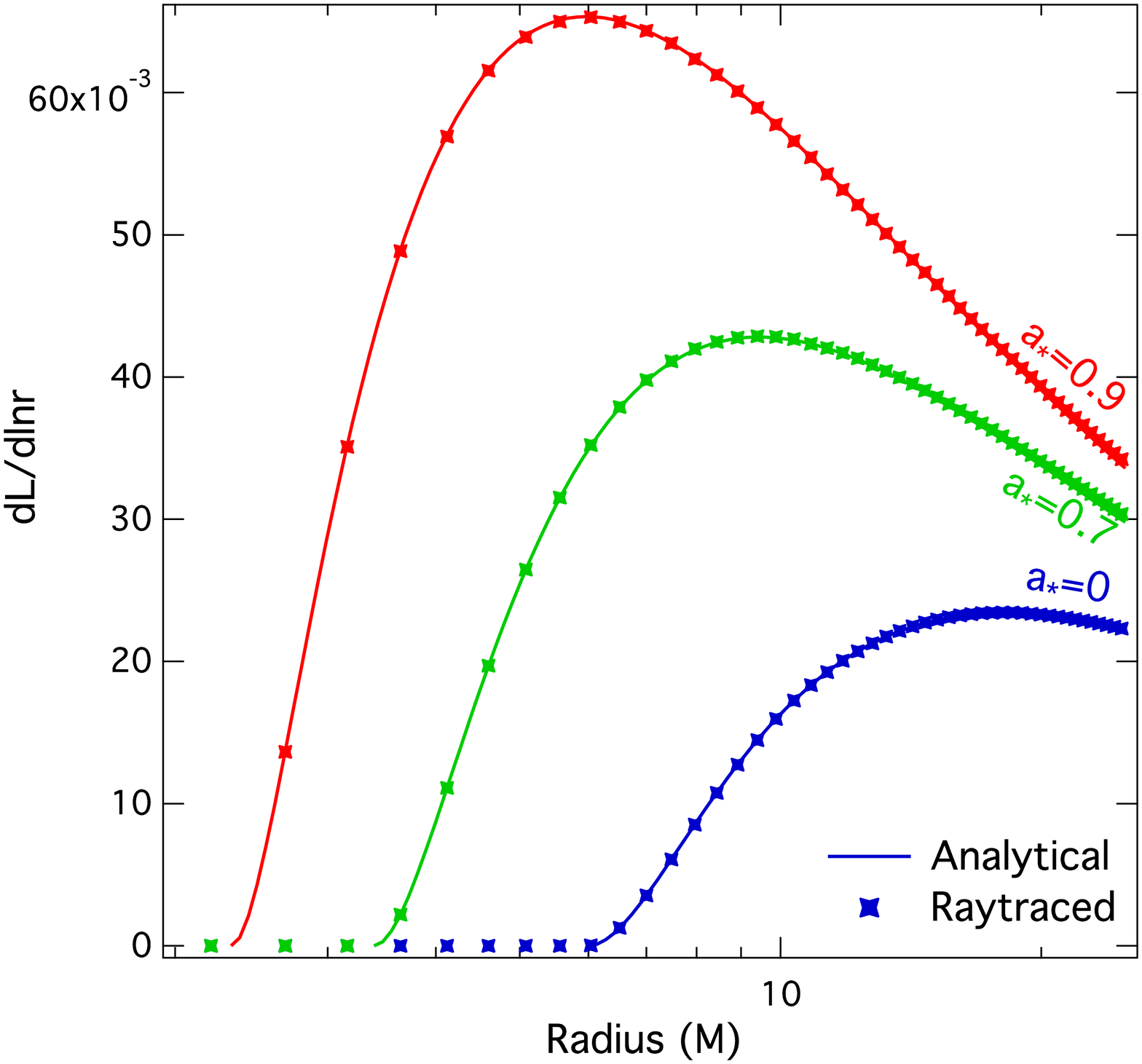}
  \caption{\emph{Left:}  The fraction of energy emitted at a radius $r$ which escapes to infinity ($f_{\rm esc}$), returns to the midplane ($f_{\rm ret}$), or is swallowed by the black hole ($f_{\rm swl}$) for three disks with Novikov-Thorne orbital velocity profiles.  The lines terminate at the photon orbit, where the orbital velocity is no longer defined, for each spin.  The locations of the innermost stable circular orbits are marked with a circle. \emph{Right:} Luminosity profiles of Novikov-Thorne disks, computed from the flux profile at the surface both analytically (solid lines) and by raytracing (points).  }
  \label{fig:NT}
\end{figure}

One additional correction is needed before we can expect our raytraced $dL/dr$ to match the analytical profile.  The analytical profile takes the flux produced by the disk and transforms it for an observer at infinity; it does not account for the fact that some photons never reach infinity but are rather swallowed by the black hole or fall back onto the disk.  These photons will be ``missing" from our raytracing solution. To account for this, we find the fraction of energy that escapes to infinity by running the raytracing code in the reverse direction, starting on the disk and following the paths of photons outward.  

Beginning in the rest frame of the fluid at a radius $r$, we distribute rays at equal intervals in $\Delta \phi_{\rm com}$ and $\Delta \theta_{\rm com}$ (fluid-frame coordinates centered at $r$) and assign them energy proportional to their solid angle $d\Omega=\sin\theta_{\rm com} d\theta_{\rm com} d\phi_{\rm com}$ times a factor of $\cos\theta_{\rm lab}$ since we are representing a surface along the midplane.  These rays are then transformed to disk coordinates in the lab frame assuming a Novikov-Thorne orbital velocity profile and integrated outward along null geodesics.

Raytracing is halted when one of three conditions is met: (i) the ray reaches the event horizon, (ii) the ray passes through the midplane, or (iii) the ray reaches some large radius and escapes.  We assign the energy carried by the rays to these three bins and divide by the total energy to get the three fractions $f_{\rm swl}(r)$, $f_{\rm ret}(r)$, and $f_{\rm esc}(r)$ which describe the fraction of energy produced at a radius $r$ which is swallowed by the black hole, returns to the disk midplane, or escapes to infinity respectively.  These fractions can be seen for disks with spin parameters $a_* = 0, 0.7,$ and 0.9 (the same spins as our GRMHD simulations) in Figure \ref{fig:NT}.  

Rays which meet the third condition and escape to infinity are the only ones accessible by the reverse-raytracing method.  Thus, we expect that the analytical and raytraced derivatives $dL/dr$ should differ by a factor of $f_{\rm esc}(r)$ so that
\begin{equation} \frac{dL}{dr} \propto \frac{1}{f_{esc}(r)} \sum\limits_{\theta_i} \frac{dF}{dr}(r, \theta_i)~ \sin\theta_i. \end{equation} 
This quantity is plotted in Figure \ref{fig:NT}, along with the analytical $dL/dr$.  A comparison of the two shows very good agreement, which demonstrates that our method captures the emission produced at the disk surface with excellent accuracy.

\section{Derivation of Redshift in Kerr Metric}
\label{sec:redshift}

The redshift we seek to derive is the shift in frequency, 
\begin{equation} 
\label{eqn:origz}
1+z = \frac{\nu_{\rm em}}{\nu_{\rm obs}} = \frac{(g_{\mu \nu} u^\mu k^\nu)_{\rm em}}{(g_{\mu \nu} u^\mu k^\nu)_{\rm obs}}, 
\end{equation}
for a photon with wave vector $k^\mu$ between an observer corotating with the disk with some circular four-velocity $u^{\mu}_{\rm em} = (u^t, 0, 0, u^{\phi})$, and a distant observer with $u^{\mu}_{\rm obs} = (1, 0, 0, 0)$.  For a spinning black hole, the metric tensor $g_{\mu \nu}$ is given by the Kerr metric:
\begin{equation} 
ds^2 = -\left(1-\frac{2Mr}{\rho^2}\right) \mathrm{d}t^2 - \frac{2Mar\sin^2\theta}{\rho^2}(\mathrm{d}t \mathrm{d}\phi + \mathrm{d} \phi \mathrm{d}t) + \frac{\rho^2}{\Delta} \mathrm{d}r^2 + \rho^2 \mathrm{d} \theta^2 + \frac{\sin^2 \theta}{\rho^2} \left[(r^2 + a^2)^2 - a^2 \Delta \sin^2 \theta \right] \mathrm{d} \phi^2 
\end{equation}
where ($t,r,\theta,\phi$) are the Boyer-Lindquist coordinates, $a = J/M$ is the angular momentum per unit mass of the black hole, $M$ is the black hole mass, and $\Delta$ and $\rho$ are given by
\begin{equation} \Delta = r^2 - 2Mr + a^2, \end{equation}
\begin{equation} \rho^2 = r^2 + a^2 \cos^2 \theta. \end{equation}
After expanding the sums in Eqn. \ref{eqn:origz} using the Kerr metric, the redshift is
\begin{equation} 1+z = \frac{(g_{tt} u^t k^t + g_{\phi \phi} u^\phi k^\phi + g_{t \phi} u^t k^\phi + g_{\phi t} u^\phi k^t)_{\rm em}}{(g_{tt} k^t + g_{t \phi} k^\phi)_{\rm obs}}. \end{equation}
This can be simplified by noting that the Kerr spacetime has two Killing vectors, $\xi^\mu = (1,0,0,0)$ and $\eta^\mu = (0,0,0,1)$, leading to two conserved quantities
\begin{equation} \varepsilon_p = -g_{\mu \nu} k^\mu \xi^\nu = -g_{tt} k^t - g_{\phi t} k^\phi, \end{equation}
\begin{equation} l_p = g_{\mu \nu} k^\mu \eta^\nu = g_{\phi t} k^t + g_{\phi \phi} k^\phi, \end{equation}
corresponding to energy and angular momentum respectively.  Then the expression for redshift reduces to
\begin{equation}
1 + z = ~u^t - u^\phi \frac{l_p}{\varepsilon_p}
\end{equation}
where all terms now correspond to the observer at the midplane.  

Making the approximation that the trajectories of photons are not affected by the curved spacetime, the spatial wave vector $\vec{k}$ from the midplane to an observer at inclination $\theta_i$ is
\begin{equation} \vec{k} = |k|(\sin\theta_i \sin\phi~\hat{r} + \cos\theta_i~\hat{\theta} + \sin\theta_i \cos\phi~\hat{\phi}) \end{equation}
in spherical polar coordinates.  The photon four-momentum $p^\mu = \hbar k^\mu$ has the property $p_\mu p^\mu = 0$.  Thus it is also true that
\begin{equation} k_\mu k^\mu = g_{\mu \nu} k^\mu k^\nu = g_{tt} k^t k^t + g_{rr} k^r k^r + g_{\theta \theta} k^\theta k^\theta + g_{\phi \phi} k^\phi k^\phi + 2 g_{t \phi} k^t k^\phi = 0. \end{equation}
We can remove $k^t$ from this expression by substituting $k^t = -(\varepsilon_p + g_{t \phi} k^\phi)/g_{tt}$ from the energy equation; then multiplying through by $g_{tt}$ yields
\begin{equation} (\varepsilon_p + g_{t \phi} k^\phi)^2 + g_{tt} g_{rr} k^r k^r + g_{tt} g_{\theta \theta} k^\theta k^\theta + g_{tt} g_{\phi \phi} k^\phi k^\phi - 2 g_{t \phi} (\varepsilon_p + g_{t \phi} k^\phi) k^\phi = 0. \end{equation}
The wave vector components $k^\mu$ are given by $k^\mu = \vec{k} \cdot \hat{e}^\mu$, where $\hat{e}^\mu$ are the dual basis vectors for the Kerr metric.  Once the $k^\mu$ are inserted, the expression has a solution of the form $|k| = \kappa \varepsilon_p$, where $\kappa$ is a function of $r$, $\phi$, $a_*$, and $\theta_i$.
The ratio $l_p/\varepsilon_p$ is, therefore, given by
\begin{align}
\frac{l_p}{\varepsilon_p} =&~\frac{g_{\phi t} k^t + g_{\phi \phi} k^\phi}{\varepsilon_p} \\
 =&-\frac{g_{\phi t}}{g_{tt}}(1+g_{t \phi} \kappa \hat{e}^{\phi \phi} \sin\theta_i \cos\phi) + g_{\phi \phi} \kappa \hat{e}^{\phi \phi} \sin\theta_i \cos\phi \\
 =& ~\kappa \hat{e}^{\phi \phi} \sin\theta_i \cos\phi \left[ g_{\phi \phi} - \frac{(g_{t \phi})^2}{g_{tt}} \right] - \frac{g_{t \phi}}{g_{tt}},
\end{align}
and the final expression for redshift is 
\begin{align}
 1 + z =& \left[u^t + u^\phi \frac{g_{t \phi}}{g_{tt}} \right] - u^\phi \kappa \hat{e}^{\phi \phi} \sin\theta_i \cos\phi \left[ g_{\phi \phi} - \frac{(g_{t \phi})^2}{g_{tt}} \right] \\
 =& ~\frac{u_t}{g_{tt}} - u^\phi \kappa \hat{e}^{\phi \phi} \sin\theta_i \cos\phi \left[ g_{\phi \phi} - \frac{(g_{t \phi})^2}{g_{tt}} \right].
\end{align}


\bibliographystyle{apj}

\end{document}